\documentclass[11pt,a4paper,titlepage,superscriptaddress,nofootinbib]{revtex4-1}
\usepackage[latin1]{inputenc}
\usepackage{amsmath}
\usepackage{amsfonts}
\usepackage{amssymb}
\usepackage{graphicx}
\usepackage{xcolor}
\usepackage{mathtools}

\newcommand{\ak}{|\vec{k}|}

\DeclarePairedDelimiter\ket{\lvert}{\rangle}

\begin{document}

\title{ 
Nambu-Goldstone Effective Theory of Information at Quantum Criticality}

\author{Gia Dvali}
\affiliation{Arnold Sommerfeld Center, Ludwig-Maximilians-University, Theresienstr. 37, 80333 M\"unchen, Germany}
\affiliation{Max-Planck-Institut f\"ur Physik, F\"ohringer Ring 6, 80805 M\"unchen, Germany}	
\affiliation{Center for Cosmology and Particle Physics, Department of Physics, New York University, 4 Washington Place, New York, NY 10003, USA}
\author{Andre Franca}
\affiliation{Arnold Sommerfeld Center, Ludwig-Maximilians-University, Theresienstr. 37, 80333 M\"unchen, Germany}
\author{Cesar Gomez}
\affiliation{Instituto de F\'{\i}sica Te\'orica UAM-CSIC, \\Universidad Aut\'onoma de Madrid, Cantoblanco, 28049 Madrid, Spain}
\author{Nico Wintergerst}
\affiliation{The Oskar Klein Centre for Cosmoparticle Physics, Department of Physics, Stockholm University, AlbaNova, 106 91 Stockholm, Sweden}

\date{\today}

 \begin{abstract}   
 
      We establish a fundamental connection between quantum criticality of a many-body system, such as Bose-Einstein condensates, and its capacity of information-storage and processing.  For deriving the effective theory of modes in the vicinity of the quantum critical  
  point we develop a new method by mapping a Bose-Einstein condensate of $N$-particles onto a sigma model 
 with a continuous global (pseudo)symmetry that mixes bosons of different momenta. The Bogolyubov modes of the condensate are mapped onto the Goldstone modes of the sigma model, which become gapless at the critical point. 
 These gapless Goldstone modes are the quantum carriers of information and entropy.  
   Analyzing their effective theory, we 
 observe the information-processing properties strikingly similar to the ones predicted by the black hole portrait. The energy cost per qubit of information-storage vanishes in the large-$N$ limit and the total 
 information-storage capacity increases with $N$ either exponentially or as a power law.  The longevity of information-storage also increases with $N$, whereas the scrambling time in the over-critical regime is 
 controlled by the Lyapunov exponent and scales logarithmically with $N$.  This connection reveals that the origin of  black hole information
storage lies in the quantum criticality of the graviton Bose-gas, and that much simpler systems that can be manufactured in table-top experiments can exhibit very similar information-processing dynamics.  
\end{abstract}
    
\maketitle

 \section{Introduction}     

  Perhaps the most mysterious thing about black holes is their capacity to store and process information.   
   Already examining the behavior of black hole entropy \cite{Bekenstein} in the classical limit, one arrives at the following puzzle \cite{QC}: according to the Bekenstein formula, the black hole entropy scales as 
   $S \sim \frac{R^2}{L_P^2}$, where $R$ is the gravitational radius and $L_P$ is the Planck length, which in terms of 
   Newton's  coupling $G_N$ and the Planck constant $\hbar$ is defined as $L_P^2 \, \equiv \hbar G_N$. 
    This expression diverges in the classical limit, if $R$ and $G_N$ are kept fixed. This is because $L_P\rightarrow 0$ in this limit. 
    Thus, a quantum physicist, to whom the classical physics is defined as a limit $\hbar \rightarrow 0$, concludes that the entropy (and thus the information-storage capacity) of a classical black hole of a finite mass is infinite.  On the other hand, a classical observer would seem to have a very different perspective. 
    
   A classical physicist knows that information can only be encoded in the black hole in features that are possible to re-arrange.
But it is well known that classical black holes are essentially featureless. They are uniquely characterized by their mass, charge and the angular momentum.  Consequently, the classical observer would conclude that black holes should carry zero entropy. Thus, it seems that the two observers must disagree on what happens in the $\hbar = 0$ limit.  
  How can infinity and zero be reconciled? 
    
     The above puzzle, along many others, can only be resolved within a microscopic theory, which explicitly identifies the quantum degrees of freedom that act as carriers of the black hole information. 
   We shall focus on a microscopic theory of this type offered in \cite{QN},  in which the black hole is described as a composite multi-particle state of soft gravitons at a {\it quantum critical} point \cite{QC}. 
  This system can approximately be described as a Bose-Einstein condensate (or a coherent state) of gravitons of characteristic wavelength $\sim R$ and occupation number $N \sim R^2/L_P^2$.
%
   
   According to this picture, the microscopic carriers of black hole entropy and information are collective excitations of the Bogolyubov-type \cite{QC}. The remarkable property of these modes is that around the critical point they become nearly gapless and decouple from the rest of the modes in the $N \rightarrow 0$ limit.  
   Hence, if a system has $N$ gapless Bogolyubov modes, there are expected to exist $\sim e^{N}$ nearly-degenerate 
 quantum states that can be labeled by the occupation numbers of these modes. 
  These states are crowded  
 within a $\sim 1/N$ energy gap.  Correspondingly, the entropy should scale as $\sim N$. 
 
  Such a behavior uncovers a very peculiar nature of black hole information storage and in particular addresses the above puzzle. In the classical limit  $N \rightarrow \infty$ and entropy is indeed infinite, but the energy gap closes to zero.  Despite the fact that the number of quantum states located within this gap diverges, the required time for resolving their differences also becomes infinite. This time-delay reconciles the views of the two observers.   
 The classical black hole appears as featureless for any finite interval of the observation time.   
   
%

 In the present paper we shall focus our attention on improving the understanding of the relation between quantum criticality and physics of information-processing.      
    Since, according to the black hole $N$-portrait  the underlying fundamental ingredient of black hole information-processing is a self-sustained quantum criticality of the cold graviton gas,  the natural question to ask is:  How important is the precise nature of the constituent bosons?  Of course, for recovering the standard gravitational properties of the system,  such as the emergence of the 
    Einsteinian metric in a mean field description, the spin-2 nature of the constituents is 
    crucial.  However,  one may suspect that the information-processing features that originate from the fact of 
    criticality, may not be so sensitive to the precise nature of the constituents.  If this is true, then a 
    wide class of critical systems must share some of the black hole information-processing abilities. 
    This would be a very important conclusion because of the following reasons. 
     First,  it would enable us to study black hole 
    properties on much simpler systems, both theoretically and experimentally.   Secondly, it would 
  open up a possibility of implementing the black hole information-processing methods in table-top experiments.
        
     Indeed, it was noticed already in \cite{QC} that the simplest prototype models of attractive Bose-Einstein condensates 
    exhibit some striking similarities with the black hole quantum portrait.   
    For example, the cost of the information-storage per qubit is by a factor of $1/N$ cheaper relative to the energy-cost exhibited  by usual (non-critical) quantum systems. In the latter systems, for the same amount of information storage,  one typically pays the energy price of the inverse size of the system, $\sim \frac{\hbar}{R}$.  Also, the degeneracy of states within the $1/N$ energy gap increases with $N$. 
    
     Moreover, it was shown that a sharp increase of one-particle entanglement takes place near the critical point \cite{ENT}   
     and the 
     scrambling of information becomes maximally-efficient 
\cite{scrambling}. 
     The time-dependent evolution of the critical condensate uncovers a scaling solution in which 
     the condensate is stuck at the critical point throughout the collapse \cite{NV}. This is the behavior that one would expect if the microscopic foundation of Hawking radiation were through the collapse and quantum depletion of the condensate. \footnote{Let us remark here that there is interesting evidence from the $S$-matrix that the above multi-particle picture represents the correct microscopic description of black hole physics. 
 Indeed,  the $2\rightarrow N$ scattering $S$-matrix element of two trans-Planckian gravitons into $N$ soft ones reproduces an entropy suppression factor $e^{-N}$  when the number $N$ is given by the 
quantum critical value \cite{scattering}. Similar conclusions were reached in \cite{KuhnelBo} in a different approach. 
For other aspects and related work, see \cite{other}.}

%
%
%
%
%
   
   In the present paper we would like to take one more step in understanding the universal role of quantum 
   criticality for information-processing.  To this end,  we invent a new method for deriving the effective low-energy theory of modes near the critical point.   As we shall see, the resulting effective theory  indeed describes 
   nearly-gapless modes with $1/N$-suppressed couplings. This form  immediately reveals the 
   key features of information storage, such as the low energy cost and the long timescale for retrieval as well as 
   the high speed of information scrambling in the over-critical unstable phase.     
    
      In our approach we formulate an effective description of quantum criticality in the isomorphic language of a pseudo-Goldstone phenomenon in a $U(2k+1)$-sigma-model. 
   The global $U(2k+1)$-symmetry of the sigma model corresponds to a continuous unitary transformation among the creation and annihilation operators of the bosons with different momenta $0, \pm 1, \pm 2, ...\pm |k|$.   This identification allows us to map the 
 Hamiltonian of the Bose-gas onto a potential of the $U(2k+1)$-sigma model.   In this way, the problem of diagonalization of the Hamiltonian becomes isomorphic to finding the symmetry breaking patterns in the sigma model.  The latter  is achieved by the straightforward minimization of the potential.  The resulting pattern determines the spectrum of pseudo-Goldstone excitations.  The quantum critical points of the Bose-gas correspond to the transitions between the different symmetry breaking patterns in the sigma model, where some Goldstone modes become gapless. 
  We show that this transition takes place when the particle number  $N$ and the strength of the attractive interaction $\alpha$ exactly balance each other, $\alpha N \, = \, 1$.   
  
  This approach allows to derive an effective Hamiltonian of the gapless excitations in form of the potential energy part of the Goldstone chiral Lagrangian.   This form reveals that near the critical point the Goldstones represent gapless excitations with $1/N$-suppressed self-interactions,  
     \begin{equation}
     H_{eff}  \, \simeq \,  \frac{n_{gold}^2}{N}  \, + \, ...  \,,    
     \label{EFFH}
 \end{equation}
  where $n_{gold}$ is the Goldstone number operator.     
  

    Analyzing the resulting effective theory, we  observe a striking connection with anticipated features of black hole information processing.  Namely, the energy cost per information qubit  goes to zero as a power of  $1/N$ and 
correspondingly near the critical point the time for information retrieval increases with $N$.  

     The number of states within the collapsed energy gap increases either as a power law or exponentially with $N$ depending on the nature of the attractive coupling.  For a non-derivative self interaction the increase follows a power law, whereas for a gravity-type derivatively-coupled case the degeneracy is increasing exponentially with $N$.      
  
  On the other hand, the instability of the system in the over-critical regime implies a positive Lyapunov exponent which, as shown in \cite{scrambling},  
is the engine for scrambling of information.    The logarithmic scaling of the scrambling time for black holes  was originally conjectured in \cite{preskill}.    The observation that critical Bose-Einstein systems are fast scramblers provides the microscopic understanding of this phenomenon in terms of a Lyapunov exponent   
and the high density of states near the quantum critical point \cite{scrambling}.

 In summary, the effective theory of information at quantum critical point developed in the present paper confirms, in accordance with \cite{QC, ENT, scrambling},  that critical multi-boson systems are indeed  efficient storers and processors of information.      
    
    Finally, in identifying the information carriers for black hole entropy, we distinguish two distinct sources of quantum degeneracy. Since degenerate states are labeled by occupation numbers of gapless Goldstone modes, 
    we can create degenerate states either by increasing the occupation numbers of a small subset of Goldstone modes,  or by diversifying the population 
  among many different Goldstone modes, while keeping individual occupation numbers small. 
  The crucial difference among the two types of states is that the information stored in 
  un-diversified large occupation numbers can in principle be resolved in interference experiments 
  even in the classical limit.  Irrespectively whether contributions from such states must be counted into black hole entropy, their existence gives a very interesting possibility of detecting the quantum sub-structure in macroscopic 
  experiments.

\section{ Understanding the Critical Point in Terms of a Pseudo-Goldstone Phenomenon} \label{sec:critgs}

 We start with the system that  was used in \cite{QC} as the simplest prototype model for  exploring the  
connection between black hole information processing and quantum criticality. It is a 
gas of cold bosons with a delta-function-type attractive interaction. Such systems are well known 
and have in particular been studied in the context of cold atoms \cite{atoms}.  
We consider a Hamiltonian of the following form,     
   \begin{equation}
 {\mathcal H} \, = \, \int d^dx \, \psi^{+} \frac{- \hbar^2 \Delta}{ 2m} \psi \, - \, g \hbar  \ 
  \int d^dx \, \psi^{+}\psi^{+} \,\psi \psi \,, 
\label{Hnonderivative} 
\end{equation} 
where $\psi \, = \, \sum_{\vec{k}} \frac{1} {\sqrt{V}} {\rm e}^{i  {\vec{k} \over R} \vec{x}} \, a_k$, 
$V = R^d$ is the $d$-dimensional volume and $\vec{k}$ is the $d$-dimensional wave-number vector.  
$a_{\vec{k}}^\dagger, a_{\vec{k}}$ are creation and annihilation operators of bosons of momentum 
${\vec{k}}$. They satisfy the usual commutation relation $[a_{\vec{k}}, a_{\vec{k'}}^\dagger] = \delta_{\vec{k}\vec{k'}}$. 
 $g$ is the coupling. Rescaling the Hamiltonian, we can write 
 $ {\mathcal H} \, \equiv \, {\hbar^2 \over 2R^2 m} \, H$.
 Introducing a notation $\alpha \, \equiv \left ({g \over V R} \right) {2Rm \over \hbar} $ and taking 
 $d=1$ we arrive at the following Hamiltonian,
 \begin{equation}
     H \, = \,  \sum_{k}  k^2 \, a_k^\dagger a_k  \, - \, {\alpha \over 4} \, \sum_{k_1+ k_2-k_3-k_4 = 0} 
   a_{k_1}^\dagger a^\dagger _{k_2}  a_{k_3}a_{k_4} \,.   
     \label{Hamilton}
 \end{equation}
 We restrict ourselves to three levels, $k=0,\pm1$.  Legitimacy of this approximation shall be justified later. 
 
  Let us define a triplet operator $a_{i} \equiv (a_{-1},a_0,a_1)$.
 Correspondingly,  we define the number operators 
 $n_{i} \equiv a_{i}^\dagger a_{i}$.  We treat the operator $a_{i}$ as a triplet under a 
 global symmetry group $SU(3)$, 
 \begin{equation} 
  a_{i} \rightarrow a_{i}' \, = \, U_{i}^{j} a_{j} \, ,         
 \label{tranform}
 \end{equation}
 where $U_{i}^{j}$ is a unitary transformation matrix which keeps invariant the 
 total number operator,  
    \begin{equation}
      n\, \equiv \,  \sum_{i}  n_{i} \,.   
     \label{number}
 \end{equation}
  Thus,  in any state with non-zero particle occupation number, 
   \begin{equation}
      \langle n \rangle \, =  \, N\,,    
     \label{numbervev}
 \end{equation} 
  the $SU(3)$-symmetry is spontaneously broken and the order parameter of breaking is $N$. 
     This fact allows us to map the dynamics of 
 the quantum phase transition in a Bose-Einstein condensate  described by the Hamiltonian (\ref{Hamilton}) 
 onto the pseudo-Goldstone phenomenon in an $SU(3)$ sigma model.   
  The Hamiltonian in terms of the triplet $a_{i}$ can be written as 
 \begin{equation}
     H \, = \,  \sum_{i=-1,1}\,  n_{i} \, -  \, {\alpha \over 2} \, (a_1^\dagger a^\dagger _{-1}  a_0a_0 \, + \, 
a_0^\dagger a^\dagger _0  a_1a_{-1})  \, + \, {\alpha \over 4} \sum_{i=-1}^1 \, n_{i}^2 \, + \, 
 H_{SU(3)} \,.   
     \label{Hamilton1}
 \end{equation}
 where,  $H_{SU(3)}$ is the $SU(3)$-invariant part of the Hamiltonian and has the following form,  
  \begin{equation}
    H_{SU(3)} \, = \,  - \,  {\alpha \over 2} \, n^2 \, + \,  {\alpha \over 4} \, n \, + \, \mu \,  (n \, - \, N) \,.
     \label{HamiltonSU3}
 \end{equation}
Here, $\mu$ is the Lagrange multiplier that fixes the total particle number.  
 Let us first investigate the ground-state of the  $SU(3)$-symmetric Hamiltonian.  Minimization with respect 
 to $n_{i}$ gives the following equations, 
   \begin{equation}
    {\partial H_{SU(3)} \over \partial n_{i} } \, = \,  - \, \alpha  \, n \, + \,  {\alpha \over 4}  + \, \mu \, = \, 0\,     
     \label{eq1}
 \end{equation}
  and 
    \begin{equation}
    {\partial H_{SU(3)} \over \partial \mu} \, = \, n \, -  \, N \, = \, 0\, \, .    
     \label{eq2}
 \end{equation}
  The vacuum is achieved at $n = N$ and $\mu \, = \, \alpha(N-1/4)$.  The $SU(3)$-symmetry is 
  spontaneously broken down to $SU(2)$ and there is a doublet of  Nambu-Goldstone bosons\footnote{When counting the number of physical Goldstone modes, we have to take into account that only different rays in the Hilbert space are independent. A breaking $SU(N) \to SU(N-1)$ can thus be understood as the breaking of $SO(N) \to SO(N-1)$ on the Hilbert space and gives rise to $N - 1$ Goldstone modes.}. For example, choosing the vacuum at $n_1=n_{-1}=0, \, n_3 \, = \, N$ and ignoring ${\cal O}(1)$-corrections to the zero mode occupation, the  Goldstone doublet is $(a_1,a_{-1})$.   The existence of the Goldstone boson is the manifestation of the indifference of the system with respect to redistributing occupation numbers
  among the different levels.  
  
  Now,  the addition of the first three terms in (\ref{Hamilton1}) breaks the 
 $SU(3)$ symmetry  explicitly and lifts the vacuum degeneracy. 
  Let us investigate the effect of these terms on the state $n_{i}  \, = \, (0,N,0)$.  Notice, this VEV continues to 
  be the extremum of the Hamiltonian even after the addition of the explicit-breaking terms, but the value 
  of  $\mu$ is now  shifted to  $\mu \, = \, {\alpha \over 2} (N-1/2)$.
   The would-be 
  Goldstone modes now acquire a non-trivial mass-matrix,  
  \begin{equation}
  \label{eq:massmat}
   \begin{pmatrix}
     1\,- \, {\alpha \over 2}N \,, & -\, {\alpha \over 2}N  \\
     - \, {\alpha \over 2}N \,, & 1\, - \,{\alpha \over 2} N
\end{pmatrix}
\end{equation}
This matrix is diagonalized by states   $a_\text{light} \, \equiv \, {1 \over \sqrt{2}} ( a_1 + a_{-1}^\dagger )$ and 
 $a_\text{heavy} \, \equiv \, {1 \over \sqrt{2}} ( a_1 - a_{-1}^\dagger )$, with 
 eigenvalues $1$ and $1\, - \,\alpha N$ respectively. In this basis the 
 Hamiltonian becomes,  
 \begin{equation}
   \begin{pmatrix}
     1 \,, & 0 \\
     0 \,, & 1\, - \,\alpha N \,.
\end{pmatrix}
\label{eq:massmatrix}
\end{equation}  
Thus,  a would be Goldstone doublet is now split into a light mode and its  heavy mode partner. 
Since the explicit breaking terms leave unbroken the $U(1)$ subgroup of $SU(3)$, corresponding to the 
generator $Q \equiv diag (1,0,-1)$,  the mass eigenstates $a_\text{light}$ and  $a_\text{heavy}$ are also  $U(1)$-charge eigenstates.  They carry charges equal to $+1$ and $-1$ respectively. In the usual language, the charge $Q$ is a momentum 
operator.
 
Let us at this point highlight the relation between the Goldstone modes $a_\text{light}$ and $a_\text{heavy}$ and the usual Bogolyubov eigenmodes of the mass matrix \eqref{eq:massmat}. The latter are obtained by diagonalizing the mass matrix through a canonical transformation; in our setup, they are given by
\begin{equation}
b_{\pm 1} = u_1 a_{\pm 1} - v_1 a_{\mp 1}^\dagger\,,
\end{equation}
with
\begin{equation}
u_1 = \frac{1 + \sqrt{1 - \alpha N}}{2(1-\alpha N)^{1/4}}\,,\hspace{2em}v_1 = \frac{1 - \sqrt{1 - \alpha N}}{2(1-\alpha N)^{1/4}}\,.
\end{equation}
In this description, the Hamiltonian has two degenerate eigenvalues \begin{equation}
\epsilon_{1,-1} = \sqrt{1-\alpha N} \,.
\label{eq:bogen}
\end{equation}
The modes $a_\text{light}$ and $a_\text{heavy}$ relate to the Bogolyubov modes through
\begin{align}
a_\text{light} &= \frac{1}{\sqrt{2}}(u_1+v_1)\left(b_1 + b_{-1}^\dagger\right) = \frac{1}{\sqrt{2} (1-\alpha N)^{1/4}}\left(b_1 + b_{-1}^\dagger\right)\,,\\
a_\text{heavy} &= \frac{1}{\sqrt{2}}(u_1-v_1)\left(b_1 - b_{-1}^\dagger\right) = \frac{(1-\alpha N)^{1/4}}{\sqrt{2}}\left(b_1 - b_{-1}^\dagger\right)\,.
\end{align}
The Goldstone mode thus encodes a light direction in configuration space, reachable through occupying both Bogolyubov modes simultaneously. Note that a state with Bogolyubov occupation number of ${\cal O}(1)$ corresponds to a Goldstone configuration with $a_\text{light}^\dagger a_\text{light} \sim 1/\sqrt{1-\alpha N}$, which is responsible for the $\sqrt{1-\alpha N}$-difference in the eigenvalues. 

The heavy partner, on the other hand, corresponds to a
heavy multi-particle direction in configuration space, reachable through populating either one of the Bogolyubov modes with ${\cal O}\left(1/\sqrt{1 - \alpha N}\right)$ particles. In turn, this explains the $\alpha N$-independence of the second eigenvalue of \eqref{eq:massmat}.

For  $1 - \alpha N \, < \, 0$ one of the pseudo-Goldstone bosons becomes tachyonic and the vacuum is destabilized.  The physics of this instability is that for $\alpha N > 1$ the  $SU(2)$ preserving vacuum 
is no longer energetically favorable and the system flows towards  the  new ground-state. This new ground-state  can be found by minimizing the full  Hamiltonian, (\ref{Hamilton1}).
 In order to minimize this Hamiltonian, let us set $a_{i}\, = \, \sqrt{n_{i}} {\rm e}^{i\theta_{i}}$. 
  The only phase-dependent term in the Hamiltonian is the second term in (\ref{Hamilton1}), which takes the form
 \begin{equation}
 -  {\alpha \over 2} \, (a_1^\dagger a^\dagger _{-1}  a_0a_0 \, + \, a_0^\dagger a^\dagger _0  a_1 a_{-1}) \, =  - \alpha \, (\sqrt{n_1n_{-1}} n_0) {\rm cos} (\theta_1+\theta_{-1}-2\theta_0) \, .
 \label{phases}
 \end{equation}
   Since there is no other conflicting phase-dependent term  in the energy,  in the minimum we will have 
    ${\rm cos} (\theta_1+\theta_{-1}-2\theta_0) \, = \, 1$.  We can thus set without any loss  of generality,
  \begin{equation}  
   \theta_1+\theta_{-1}-2\theta_0 \, = \, 0 \, . 
   \label{phaseconstraint}
    \end{equation}
 Another simplifying observation is that in the extremum 
  $n_1$ and $n_{-1}$ must be equal.  This can be seen by extremizing the Hamiltonian with respect to $n_1$ and $n_{-1}$, 
    \begin{equation}
    {\partial H \over \partial n_{1} } \, = \,  1 \, - \, {\alpha \over 2}  \sqrt{{n_{-1} \over n_1}} n_0\, + \, {\alpha \over 2}n_1
  -  \, \alpha  \, n \, + \,  {\alpha \over 4}  + \, \mu \, = \, 0\,     
     \label{min1}
 \end{equation}
  \begin{equation}
    {\partial H \over \partial n_{-1} } \, = \,  1 \, - \, {\alpha \over 2}  \sqrt{{n_1 \over n_{-1}}} n_0\, + \, {\alpha \over 2}n_{-1}
  -  \, \alpha  \, n \, + \,  {\alpha \over 4}  + \, \mu \, = \, 0\,     
     \label{min2}
 \end{equation}
  which show that $n_1$ and $n_{-1}$ can be non-zero only together. 
 Moreover they must be equal, since they satisfy the same equation, with only one positive root. 
  This can be seen by multiplying  (\ref{min1})  and (\ref{min2}) by $n_1$ and $n_-1$ respectively. The resulting quadratic equations are identical, 
   \begin{equation}
   n_1 {\partial H \over \partial n_{1} } \, = \, {\alpha \over 2}n_1^2 \, + \, n_1(1 \, - \, \alpha N \, +  {\alpha \over 4}  + \, \mu) \, - \, {\alpha \over 2} \sqrt{n_{-1} n_1} n_0\,  = \, 0\,     
     \label{minim1}
 \end{equation}
  \begin{equation}
   n_-1 {\partial H \over \partial n_{-1} } \, = {\alpha \over 2}n_{-1}^2 \, + \, n_{-1}(1 \, - \, \alpha N \, +  {\alpha \over 4}  + \, \mu) \, - \, {\alpha \over 2} \sqrt{n_{-1} n_1} n_0\, = \, 0\,     
     \label{minim2}
 \end{equation}  
  and have only one positive root. Thus, without any loss of generality we can minimize the Hamiltonian for the 
  configuration $n_{i} \, = \, (x,N-2x,x)$,  which gives 
   \begin{equation}
     H \, = \,  {7\over 2} \alpha  x^2 \,  + \, 2x(1- \alpha N) \, - \,  {\alpha \over 4}N\left(N-1\right) \,.   
     \label{HamiltonX}
 \end{equation}
    Since $x$ is positive definite, for $\alpha N <1$, the minimum is achieved at  $x=0$. For $\alpha N > 1$, the minimum is at 
   \begin{equation}
   x \, = \, {2 \over 7\alpha} (\alpha N -1)  \, . 
   \label{generalmin}
   \end{equation} 
Not surprisingly, this corresponds to the first Fourier modes of the bright soliton solution to the Gross-Pitaevskii equation in the overcritical regime \cite{atoms}.
   The energy of the ground-state is given by
  \begin{equation}
     H_{min} \, = \, - \, {2 \over  7\alpha } \,(\alpha N\, -\,1)^2 \, - \,  {\alpha \over 4}N(N-1) \,,    
     \label{gsenergy}
 \end{equation}
where the first term only exists for $\alpha N > 1$.   

  We can now understand the critical phenomenon in terms of this Goldstone-mode. 
 The large occupation number serves as an order parameter for a spontaneous breaking of
  a global $SU(3)$-symmetry. This symmetry corresponds to redistribution of the particle occupation numbers 
  between the different momentum states, without changing their total number. 
   This symmetry is only approximate, and normally the would-be Goldstone mode has the mass of the order of the 
   first momentum level (that is order one in our units).  However,  the phase transition corresponds to the point 
   where the particle number distribution changes.  This necessarily implies that the corresponding pseudo-Goldstone mode must become massless at this point. 
   
   Notice that since $x$ parameterizes the occupation number of (pseudo-)Goldstones,  $x \,  
  = n_{gold}  =  a_{gold}^\dagger a_{gold}$,
    the  Hamiltonian (\ref{HamiltonX}) essentially represents  the effective action for this mode, 
    \begin{equation}
     H_{Gold}  \, = \,  (n_{gold})^2 \, \alpha_{gold} \,  + \, n_{gold} m^2_{gold}  - \,  {\alpha \over 4}N\left(N-1\right) \,.    
     \label{Goldeff}
 \end{equation}
 Thus, the mass of the pseudo-Goldstone is given by $m^2_{gold} \, \equiv \, 2(1- \alpha N)$ and the self-coupling 
 is given by $\alpha_{gold} \equiv   \alpha {7\over 2}$. 
   At the critical point, the Goldstone mass term vanishes and the theory is described by a gapless mode with 
    a self-interaction strength given by $\alpha_{gold} \,  = \, {7/2N}$,  
         \begin{equation}
     H_{Gold}  \, = \,  (n_{gold})^2  {1\over N}  {7\over 2} \, - \,  {1 \over 4}\, N \,.   
     \label{Goldeffcritical!}        
 \end{equation}

 
Of course, the phenomena uncovered in the Goldstone formulation are in one-to-one correspondence to those that are seen in a mean-field and Bogolyubov description of the model. In fact, as hinted before, the minima of the Hamiltonian  Eq.\eqref{HamiltonX} correspond to the $k = -1,0,1$ modes of the solutions to the mean-field Gross-Pitaevskii equation: for $\alpha N < 1$ we found the homogenous condensate, while the solution for $\alpha N > 1$ corresponds to a localized bright soliton. 
\begin{figure}
	\includegraphics[width=0.49\linewidth]{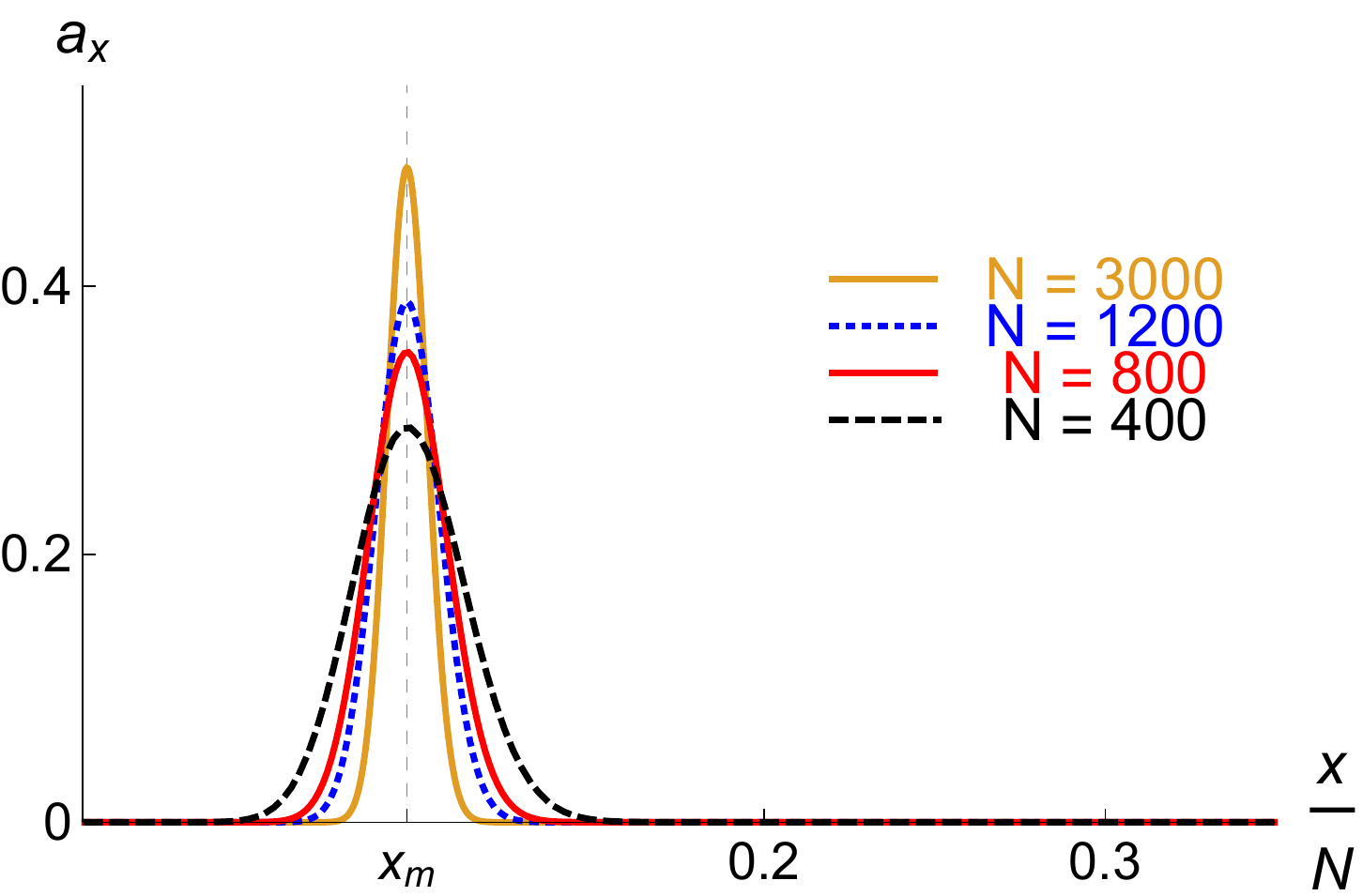}
	\includegraphics[width=0.49\linewidth]{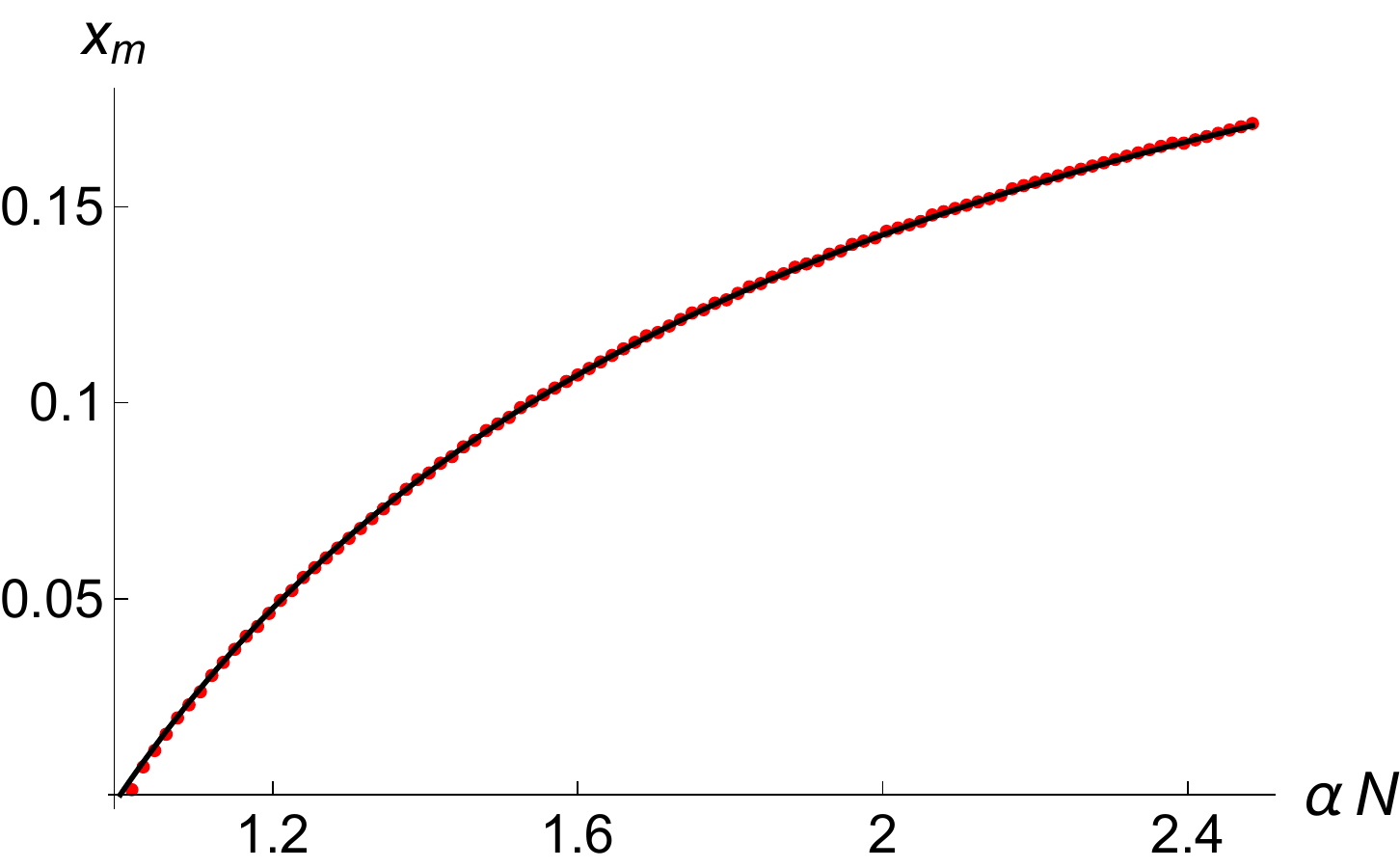}
	\caption{{\bf(a)} Ground state for $\alpha N=1.5$ at various values of $N$, using the parametrization $ \ket{GS} = \sum_{x=0}^{N/2} a_x \ket{x,N-2x,x} $. $a_x$ are normalized according to $\sum_{x=0}^{N/2} a_x^2 = N$, since as $N\rightarrow \infty$, $\int_{0}^{1/2} a(Ny)^2 \, dy = 1$ and thus it's possible to compare them at different $N$. As $N$ increases, the distribution becomes increasingly more peaked at $x_m$. {\bf(b)} $x_m $ for $N=800$ particles compared with the analytic prediction $x_m = \frac{2}{7\alpha N} (\alpha N -1) $. }
	\label{fig:gsfig}
\end{figure}
In this context, one should also keep in mind that the solution $(x, N-2x, x)$ only agrees with the exact quantum mechanical ground state in the limit $N \to \infty$. At any finite $N$, the ground state is a smeared distribution centered around $(x, N-2x, x)$ with $x$ given by \eqref{generalmin}, as can be seen in Fig. \eqref{fig:gsfig}. In particular, the ground state at the critical point is characterized by strong entanglement and is therefore not well described by a mean field. 
This, however, will not alter our conclusion on state-counting and information processing to be performed in the following sections.

We reemphasize that in the Goldstone language, the critical point amounts to destabilization of the $SU(2)$-invariant vacuum.  
   With the pseudo-Goldstone method we have traded the diagonalization of the Hamiltonian with the minimization 
   procedure.  Substituting the small deformations of the order parameter by the Goldstone mode allowed us to derive 
the effective action for the latter.   This is similar to deriving an effective Hamiltonian of a phonon field in a background external magnetic field, which breaks the rotational symmetry of a ferromagnet.   The analogous role in our treatment is assumed by the terms in the Hamiltonian that explicitly break 
 $SU(3)$-symmetry. \footnote{Notice also a curious analogy with the sigma-model of large-$N$-color QCD. 
  In this case our $N$ would be mapped on the number of colors, whereas the levels $k=\pm,0$ to quark 
  flavors. The pseudo-Goldstone boson then is analogous to pion, which also has $1/N$ suppressed
  self-coupling. }

We conclude this section by pointing out that one can continue an analytic treatment of the system into the solitonic regime by performing an $x$-dependent canonical rotation on the creation and annihilation operators $a_i$. We define the condensate mode
\begin{equation}
c_\text{cond} = \sqrt{\frac{x}{N}}a_{-1} + \sqrt{1 - \frac{2x}{N}}a_{0} + \sqrt{\frac{x}{N}}a_{1}\,,
\end{equation}
and the corresponding orthogonal modes
\begin{equation}
c_{\pm 1} = \frac{1}{2}\left[\left(\pm 1+\sqrt{1-\frac{2x}{n}}\right)a_{-1} -2\sqrt{\frac{x}{N}}a_0 + \frac{1}{2}\left(\mp 1+\sqrt{1-\frac{2x}{n}}\right)a_{-1}\right]\,.
\end{equation}
The resultant Hamiltonian is now always minimized by the configuration $(c_{-1},c_\text{cond},c_1) = (0,N,0)$. Expanding around this vacuum leads to a quadratic Hamiltonian which can be written in matrix form as
\begin{equation}
H_2 = \left(c_1^\dagger,c_{-1}^\dagger,c_1,c_{-1}\right)
\begin{pmatrix}
m_1 & m_2 & m_3 & m_4\\
m_2 & m_1 & m_4 & m_3\\
m_3 & m_4 & m_1 & m_2\\
m_4 & m_3 & m_2 & m_1
\end{pmatrix}
\begin{pmatrix}
c_1\\c_{-1}\\c_1^\dagger\\c_{-1}
\end{pmatrix}
\end{equation}
with
\begin{equation}
\begin{pmatrix}
m_1\\m_2\\m_3\\m_4
\end{pmatrix}
=
\begin{pmatrix}
1 - \frac{\alpha N}{2} + x \left(8 \alpha - \frac{3 + 14 \alpha x}{N}\right)\\
\frac{x}{N}\left(-1 + 3\alpha N - 7 \alpha x\right) \\
\frac{\alpha x}{2}\left(3 - 7 \frac{x}{N}\right)\\
\frac{\alpha}{2}\left(4 x  - N - \frac{7 x^2}{N}\right)
\end{pmatrix}
\end{equation}
 Canonical diagonalization again leads to two distinct eigenmodes; for $\alpha N < 1$, their energy is $\epsilon = \sqrt{1 - \alpha N}$. For $\alpha N > 1$, one of them remains exactly gapless, while the other has a frequency $\epsilon = \sqrt{1-\alpha N}\sqrt{\frac{2}{7}(4+3\alpha N)}$. The former is the mode that corresponds to translations of the soliton, whereas the other is the Bogolyubov mode that is only light in the vicinity of the critical point. In the Goldstone language, the Goldstone mode $a_\text{light}$ splits into two parts, essentially given by its real and imaginary parts. One remains gapless and generates translations of the localized ground state; the other is gapless only at the critical point and obtains a frequency $\epsilon = (1-\alpha N) \frac{2(4+3\alpha N)}{7 \alpha N}$ for $\alpha N > 1$.
 
 The appearance of the light modes can of course again be related to the breaking of $SU(3)$ generators. Keeping in mind that in the rotated description the symmetry transformations are given by $c' = S U S^{-1} c$, where $S$ is the symplectic matrix that generates the canonical transformation, we immediately see that the translation generator $Q = diag(1,0,-1)$, which commutes exactly with the Hamiltonian for all $x$, does no longer annihilate the groundstate. The generator that redistributes particle numbers, on the other hand, only commutes with $H$ at the critical point.

\section{Information Processing}  

  The above form \eqref{Goldeff} of the effective Hamiltonian displays the role of quantum criticality for information 
  storage and processing. The quantum information in the above system is encoded in the state of the Goldstone mode.  The remarkable thing about it is the low energy cost of information-qubit-storage, which is suppressed by powers
  of $1/N$ relative to the inverse size of the system.  This phenomenon is a manifestation  of quantum criticality.

%
   In the time evolution of the Goldstone state we can distinguish two sources. 
   One is the interaction governed by a quartic self-interaction Hamiltonian
   (\ref{Goldeffcritical!}). The rate of this process is suppressed by powers of  $1/N$, and correspondingly the time-scale 
   of evolution is very long.  
   
   The second source of  time evolution is the Goldstone mass term  that parametrizes the departure from quantum criticality. In the overcritical regime, this mass is imaginary and results in an exponential growth of the Goldstone occupation number.  This instability is described by a Lyapunov exponent that, as 
   shown in \cite{scrambling}, leads to the 
   generation of one-particle entanglement and potentially to the scrambling of information.
   
   \section{Numerical results of State Evolution}
   
   An exact numerical diagonalization of the Hamiltonian \eqref{Hamilton} provides a complementary analysis which is valid also at the critical point. Using the same technique as in \cite{ENT}, we can verify the above results by comparing the 
   the first Bogolyubov state as well as a lowly occupied Goldstone state to the exact eigenstates of the Hamiltonian \eqref{Hamilton}.
   Several quantities are instructive. 
   
\begin{figure}[t!]
	\centering
	\includegraphics[width=0.47\linewidth]{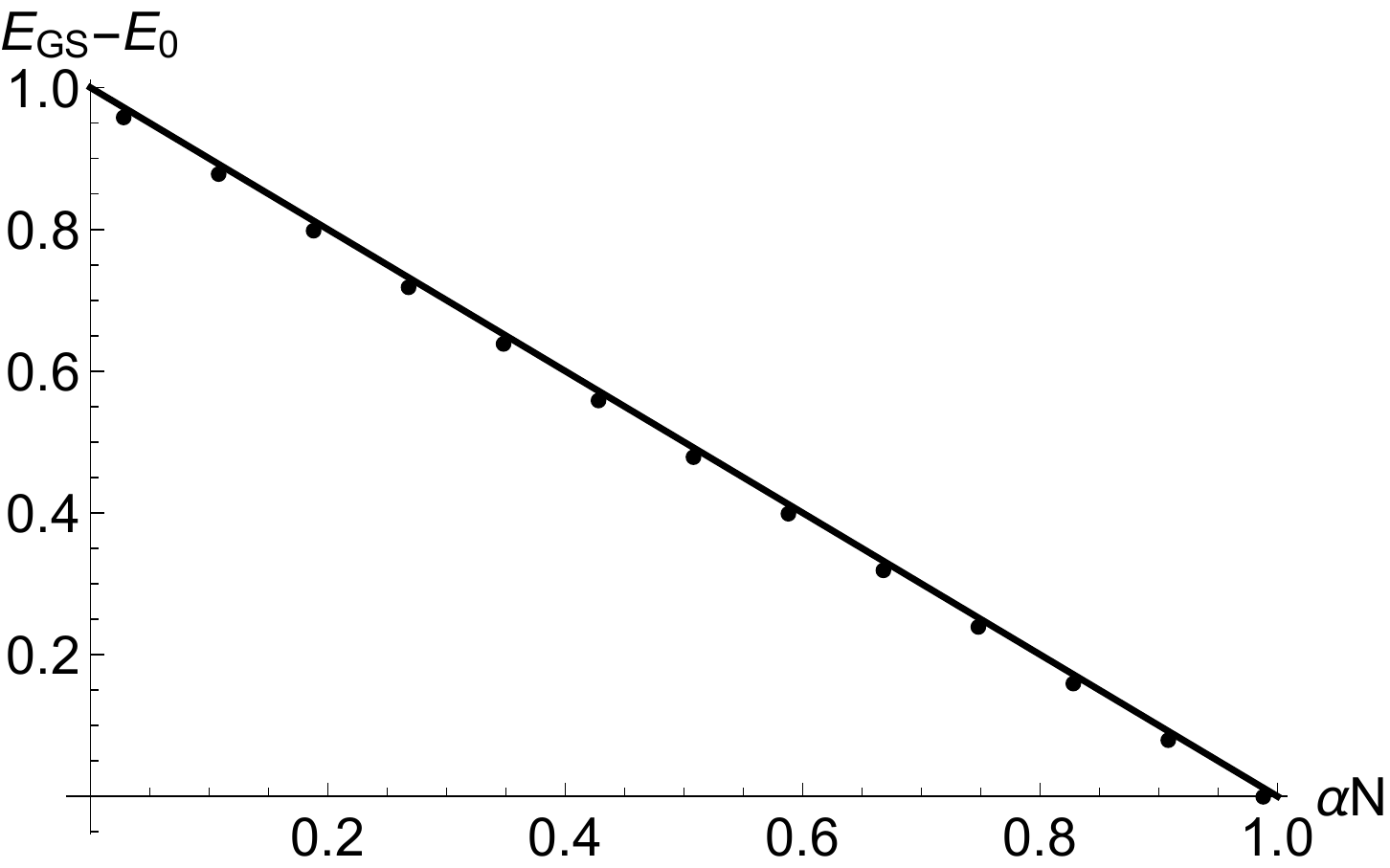}            
	\caption{Expectation value of energy in state with Goldstone occupation $\left\langle a_\text{light}^\dagger a_\text{light}\right\rangle = 1$} as a function of $\alpha N$ for $N = 500$.
	\label{fig:gsenergy}
\end{figure}

   In Fig.\ref{fig:gsenergy} we plot the expectation value of the energy in a state with Goldstone occupation $\left\langle a_\text{light}^\dagger a_\text{light}\right\rangle = 1$ as a function of $\alpha N$. The solid line corresponds to the analytic result \eqref{Goldeff}. The increase of energy around the phase transition is suppressed by $1/N$ and thus not visible.

\begin{figure}[t!]
    \centering
    \includegraphics[width=0.47\linewidth]{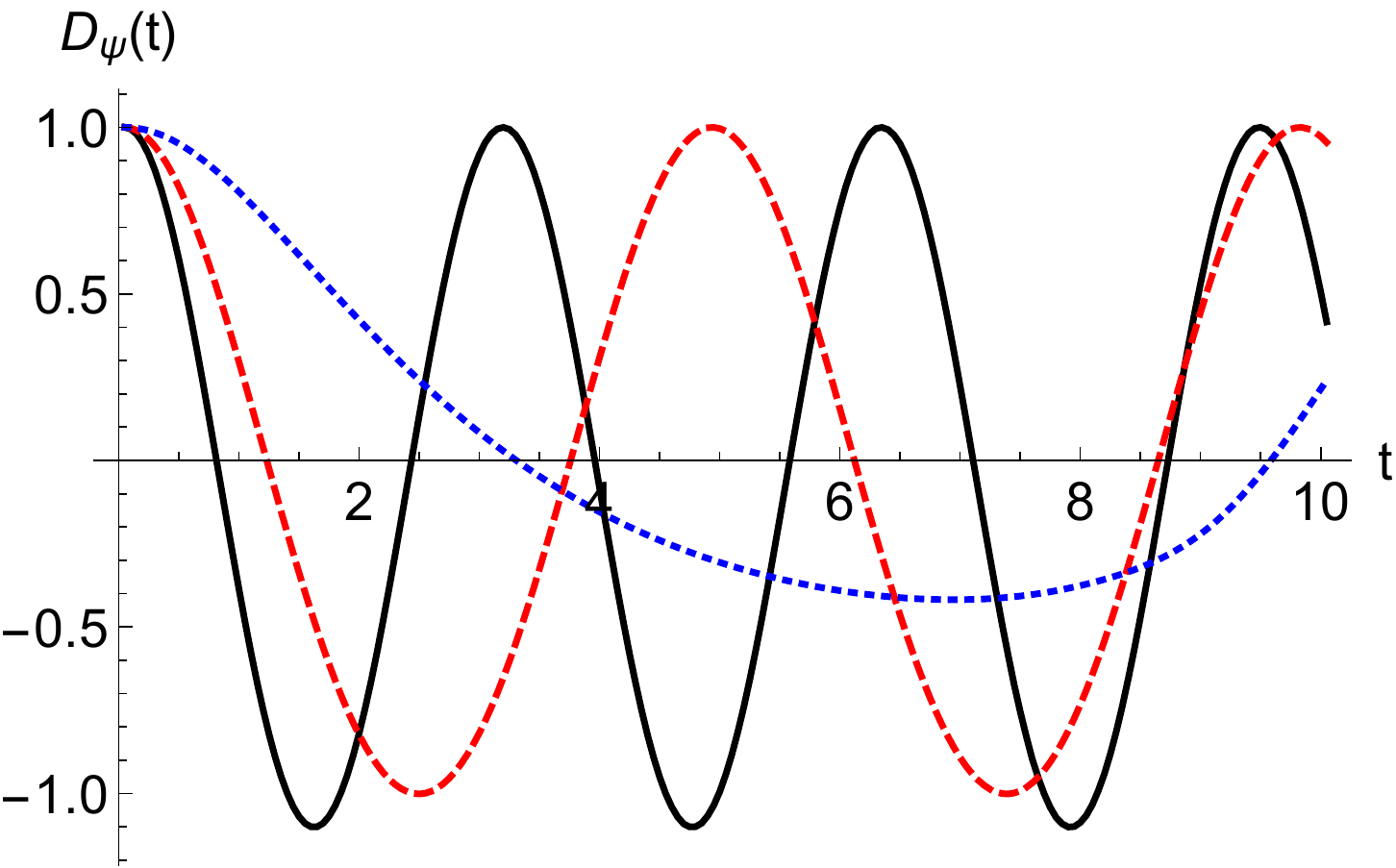}
    \hspace{0.02\linewidth}
    \includegraphics[width=0.47\linewidth]{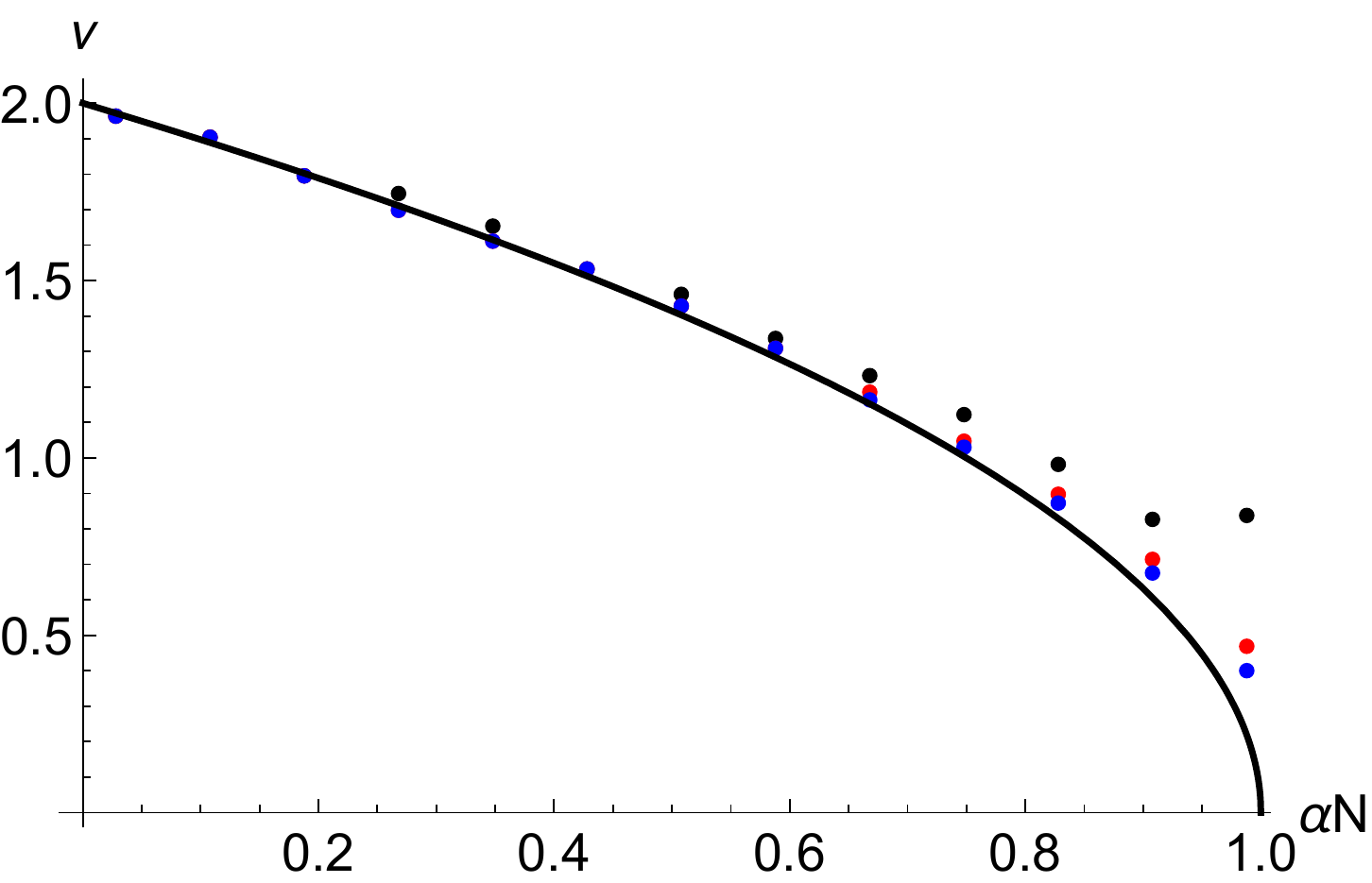}            
    \caption{{\bf (a)} Normalized exact time evolution of the time dependent part of the first Bogolyubov state for $N = 500$ and $\alpha N = 0.004$ (black, solid), $0.596$ (red, dashed) and $0.976$ (blue, dotted).
    {\bf (b)} Frequency of oscillation as a function of $\alpha N$ at various values of $N$.  The solid line corresponds to the analytic result $2\sqrt{1-\alpha N}$}
    \label{fig:time_evol}
\end{figure}

   In Fig.\ref{fig:time_evol}a we plot the exact time evolution of the Bogolyubov state $\ket{1_B} = b_1^\dagger b_{-1}^\dagger \ket{0_B}$, where $\ket{0_B}$ is the Bogolyubov ground state. The results are obtained for fixed particle number $N = 500$, while the effective coupling $\alpha N$ is varied. We observe an decrease of the frequency as $\alpha N$ approaches the critical point. 
   To illustrate this point better, we plot the frequency of oscillations versus $\alpha N$ for $N = 100,\,300$ and $500$ in Fig \ref{fig:time_evol}b. We observe the square root behavior \eqref{eq:bogen}, with the frequency scaling with $\alpha N$ as $\sqrt{1 - \alpha N}+{\cal O}(1/N)$.

  \section{Occupation and Stability of Higher Modes} 
  
    We can now verify that our starting assumption about the stability and a negligible occupation number 
    of higher momentum modes is valid 
    as long as ${x \over N} < 1$.    As we shall see,  the occupation numbers of modes with momentum number 
    $k$ are suppressed  by a factor of $(x/N)^{|k|-1}$.  This is a manifestation of the fact that 
    the global $SU(2|k| + 1)$-symmetry, corresponding to redistribution of occupation numbers 
    among the fist $k$ levels, is explicitly broken by the momentum term in the Hamiltonian (\ref{Hamilton}). 
    We can think of this breaking as the decoupling limit of a {\it spontaneous}  breaking by a 
    spurion order parameter, $\Sigma$, in the adjoint representation of the $SU(2|k|+1)$ group, with the expectation value 
    $\Sigma \, = \, {\rm diag} (-|k|, \ldots ,-1,0,1,\ldots,|k|)$.\footnote{This expectation value leaves invariant 
    the $U(1)$-subgroup with the generator $Q$  which is proportional to  $\Sigma$. As in the case of 
    $SU(3)$, this charge represents a momentum operator.}   
    Obviously, the kinetic term in the Hamiltonian  (\ref{Hamilton}), 
   can then simply be written as an  $SU(2|k|+1)$-invariant product, $a^\dagger \, \Sigma^\dagger \Sigma \, a$.  
   
      The spurion $\Sigma^\dagger \Sigma$ can be thought of as an expectation value of the number operator of 
     a "hidden" Bose-gas of some $\sigma_{i}$-particles, interacting with $a_{i}$ by an infinitely weak coupling $\kappa$, s.t. $\Sigma^{+} \Sigma \, = \, \kappa \sum \langle \sigma_{i}^\dagger  \sigma_{i} \rangle$.  In that case, the momentum operator is replaced by a peculiar $SU(2|k|+1)$ invariant interaction 
     term and  both occupation numbers spontaneously break one and the same $SU(2|k|+1)$-symmetry. 
   The  Goldstones bosons are superpositions of $a$-s and $\sigma$-s.   The orthogonal combinations are 
   pseudo-Goldstones.   
   
    Now, let us take the limit $\kappa \rightarrow 0$, while simultaneously taking the occupation numbers of 
    $\sigma$-particles to infinity in such a way that the products $ \kappa \langle \sigma_{i}^\dagger  \sigma_{i} \rangle  = i^2$ are kept fixed for all $i \, = \, -1,1, ...$.  We arrive at the situation in which 
  in the $a$-particle sector, the entire information about the breaking is summed up in the expectation value of the spurion  $\Sigma$.  Correspondingly, all the Goldstones in the $a$-sector become pseudo-Goldstones.  With increasing momentum level $|k|$,  the masses of pseudo-Goldstones increase whereas their occupation numbers rapidly diminish.  
  
      In order to see this, let us compute the occupation numbers of the $k = \pm 2$ modes. The corresponding annihilation operators 
    for $k= + 2$ and $k=-2$  are $a_{2}$ and $a_{-2}$, respectively. 
    The bilinear mass matrix of  these modes  
    expanded about the state  $n_{i} \, = \, (0,x,N-2x,x,0)$, has the form 
     \begin{equation}
   \begin{pmatrix}
     4\,- \, \alpha N \,, & -\, {\alpha \over 2}N  \\
     - \, {\alpha \over 2}N \,, & 4\, - \,\alpha N
\end{pmatrix}
\end{equation} 
     This has two eigenvalues, $4\,- \, {1\over 2} \alpha N$ and   $4\,- \, {3\over 2} \alpha N$. 
   Both eigenstates are stable as long as $\alpha N < 8/3$. 
  
    However,  the modes $n_{-2,2}$ nevertheless can get populated due to the mixing with the lower levels. 
    The occupation number of  the lower modes acts as a source for the higher ones. 
    The occupation numbers of the higher modes can be easily obtained by 
    minimizing  the bilinear Hamltonian including the effective source terms for 
  $a_{-2,2}$. This source is  generated from the interaction term after plugging the expectation values of 
  $n_{-1,0,1}$.  The phases of the expectation values can again safely be set to zero, because 
  all the source terms are negative.  The corresponding effective Hamiltonian for $a_{2}$ and $a_{-2}$ has the form, 
 \begin{eqnarray}
     H_{2,-2} \, & = &\, {1\over 4}  \left (  (8 - 3 \alpha N ) (a_{2}^\dagger  \, + \, a_{-2}) 
 (a_2 \, + \, a_{-2}^\dagger )   \, + \,  (8 - \alpha N)  (a_{2}^\dagger  \, - \, a_{-2})(a_{2} \, - \, a_{-2}^\dagger ) \right )\, - \, \nonumber \\ 
 && - \, {3 \over 2} \alpha x \sqrt{(N-2x)} ( a_2 \, + \, a_{-2}^\dagger ) \, +  {\rm h.c.}   
     \label{bilinear45}
     \end{eqnarray}

  Minimizing this with respect to $a_{-2,2}$ we see 
  that 
\begin{equation}
  n_{-2,2} \sim x^2 /N\,.
  \label{eq:k2quad}
\end{equation}  
  Similarly, we can derive the occupation numbers of higher momentum modes. 
  In general, modes with momentum $|k|$ have occupation numbers 
  $ \sim \,  x (x/N)^{|k|-1}$, which rapidly approaches zero for $x/N \ll 1$. 
  Thus in this regime, our assumption is well justified.  In particular, this regime covers the neighborhood 
  of the phase transition, around which $x$ is small. 
  
\begin{figure}[t!]
    \centering
    \includegraphics[width=0.49\linewidth]{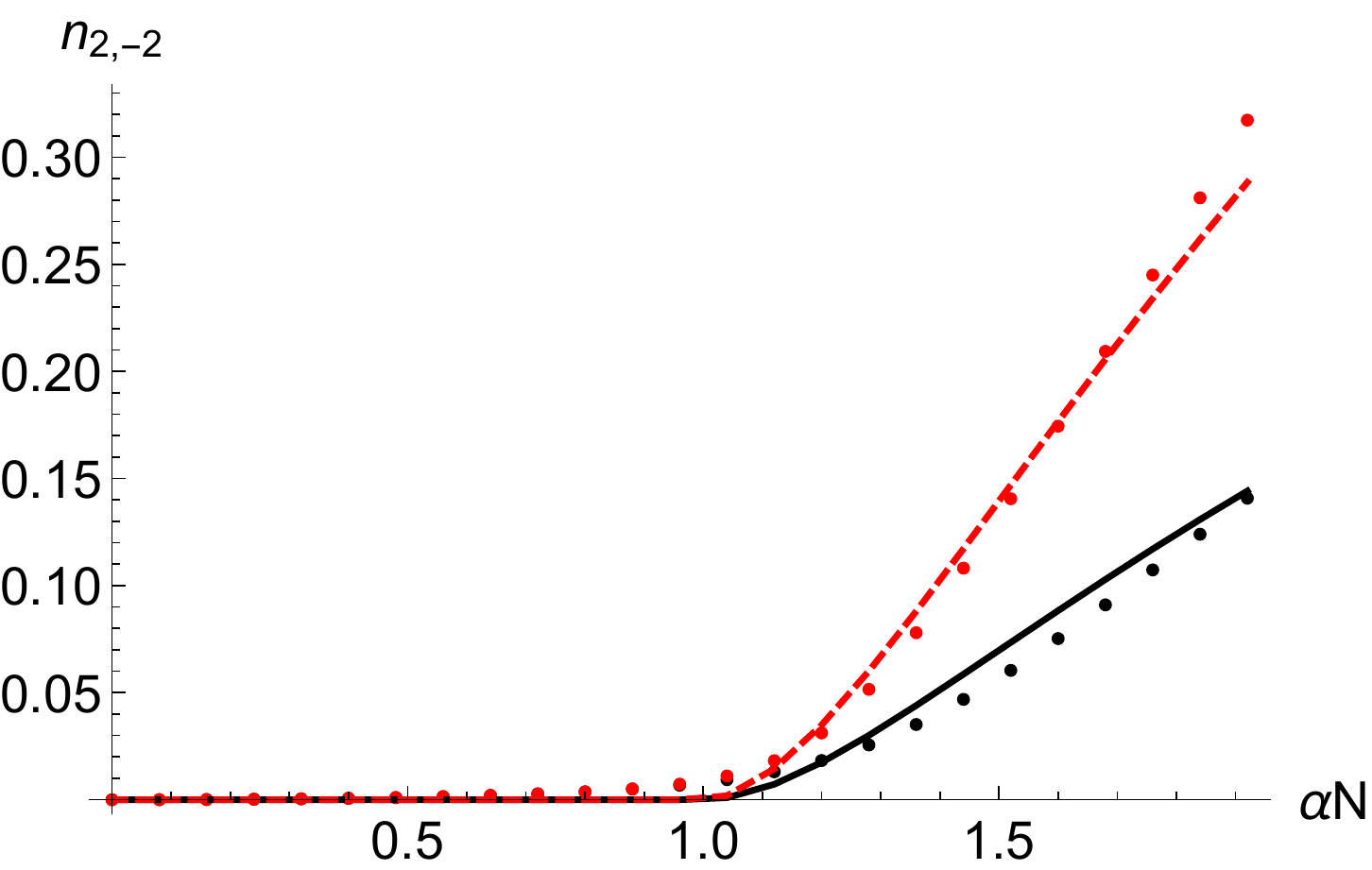}
    \caption{Contribution of the $|k|=2$ modes to the lowest lying eigenstates for $N = 10$ (black) and $N = 20$ (red). The dots correspond to the exact numerical results, while the solid line depicts the analytical behavior \eqref{eq:k2quad}.}
    \label{fig:k2}
\end{figure}

  This behavior can also be confirmed by a numerical study of the system. To this end, we consider the Hamiltonian \eqref{Hamilton} with a momentum cut-off at $|k| = 2$. An instructive quantity is the occupation number $n_{-2,2}$ of the $|k| = 2$ modes for low lying states around the phase transition. As can be seen in Fig. \ref{fig:k2}, there is no appreciable contribution of the $|k| = 2$ modes to the low lying states at least until $\alpha N \sim 2$. We can moreover confirm the quadratic behavior \eqref{eq:k2quad}.
  
 \section{Derivatively-Coupled Case} 
 
The model considered in the previous sections admits two solutions that become degenerate for $N \to \infty$ at the critical point. In $d$ dimensions, this number can be increased by a factor of $d$; this corresponds to the number of gapless Bogolyubov modes. For finite $N$, we may define an entropy for a given $N$-particle system at the critical point by counting all states with $\frac{E - E_0}{E} < 1/N$. Given the gap of the light modes, $\Delta E \sim 1/N$, we obtain $\sum_k^N {d + k - 1 \choose k} \sim N^d$ states within the accountable range of energies.  This implies an entropy $S \sim d \log{N}$.

We may attempt to reproduce an entropy that scales with $N$ like the Bekenstein-Hawking entropy of black holes, $S \sim N$, by increasing the number of pseudo-Goldstone modes at the critical point. This can be achieved, for example, by taking the coupling to be momentum-dependent.
 In order to see this,  consider a Hamiltonian, 
 \begin{equation}
 {\mathcal H} \, = \, \int d^dx \, \psi^{+} {- \hbar^2 \Delta \over 2m} \psi \, - \, L_*^{d+1} \hbar  \ 
  \int d^dx \left(\psi^{+} \vec{\nabla}\psi^{+}\right) \, \left(\psi  \vec{\nabla}  \psi \right) \,, 
\label{Hderivative} 
\end{equation} 
where $\psi \, = \, \sum_{\vec{k}} {1 \over \sqrt{V}} {\rm e}^{i  {\vec{k} \over R} \vec{x}} \, a_k$, 
$V = R^d$ is the $d$-dimensional volume and $\vec{k}$ is the $d$-dimensional wave-number vector.  
 $L_*$ is a fundamental length and sets the cutoff of the effective theory. Rescaling the Hamiltonian, we can write 
 $ {\mathcal H} \, \equiv \, {\hbar^2 \over 2R^2 m} \, H$, where
 \begin{equation}
     H \, = \,  \sum_{\vec{k}}  \vec{k}^2 \, a_k^\dagger a_k  \, - \, \alpha_0  \, \sum_{k_1+ k_2-k_3-k_4 = 0} 
  \left(\vec{k}_2\vec{k}_4 \right) a_{k_1}^\dagger a^\dagger _{k_2}  a_{k_3}a_{k_4} \,.   
     \label{Hamiltonderexp}
 \end{equation}
 and $\alpha_0 \, \equiv \left ({L_*^{d+1} \over V R} \right) {2Rm \over \hbar} $.   Let us again find the effective bilinear Hamiltonian for  $k\neq 0$ modes, about the  point $n_0 \, = \, N$ and 
$n_{k \neq 0} \, = 0$.  We obtain    
  \begin{equation}
     H \, = \,  \sum_{\vec{k}\neq 0}  \vec{k}^2 \, \left( \, 1 \,  -  \, \alpha_0 N \right ) 
   a_{\vec{k}}^\dagger a_{\vec{k}}       
     \label{bilinearcrit}
 \end{equation}
   Thus, at $\alpha_0 \, = \, 1/N$, all modes are critical and there  is the same number of massless pseudo-Goldstones as the number of 
   momentum modes.
   
   This peculiar behavior can be equivalently understood in a mean field analysis. Minimizing Eq. \eqref{Hderivative} under the constraint of fixed total particle number yields the Gross-Pitaevskii equation 
   \begin{equation}
   \label{eq:gpder}
-\frac{\Delta}{2m}\psi + \frac{L_*^{d+1}}{2}\psi^\dagger \Delta{\psi^2} = \mu \psi\,,
\end{equation}    
where $\mu$ is again the Lagrange multiplier fixing the total particle number $\int dV \psi^\dagger \psi = N$. Using the ansatz $\psi_0^{\vec{k}} = \sqrt{\frac{N}{V}} e^{i \vec{k} \vec{x}}$, Eq.\eqref{eq:gpder} becomes
\begin{equation}
\frac{k^2}{m}\left(\frac{1}{2} - \alpha_0 N\right) = \mu\,.
\end{equation}
Hence plane waves of arbitrary wavenumber solve the GP equation \eqref{eq:gpder}. By inserting $\psi_0^{\vec{k}}$ into the Hamiltonian \eqref{Hderivative}, we read off the energy of the plane wave solutions
\begin{equation}
E_k = \frac{\vec{k}^2}{2m}\left(1 - \alpha_0 N\right)\,.
\end{equation}
At the critical point, all plane wave solutions become degenerate; the corresponding modes $a_{\vec{k}}$ are the Bogolyubov modes.
   
    In order to give a lower bound on the entropy of this system for a given $N$, we have to keep in mind that for the validity of the non-relativistic treatment we should only include modes with 
   $|\vec{k}| \, < \, k_{max} \equiv mR/\hbar$. Their number is $N_{max} \, \approx \, (k_{max})^d$.  This number sets the 
 number of legitimate  pseudo-Goldstone modes. However,  not all the massless-pseudo-Goldstone modes 
 contribute the same weight into the entropy. This is because their self-couplings have  different strengths.
 Hence exciting different pseudo-Goldstone species will contribute into the Hamiltonian differently.  In order to identify which species 
   give the maximal contribution into the entropy of the system, let us consider the effective Hamiltonian for the 
   pseudo-Goldstone modes. 
   It has the following form (we ignore numerical factors OF order one), 
    \begin{equation}
     H_{Gold}  \, = \, \sum_{\vec{k}\neq 0}  \, |\vec{k}|^2 \left ( \left(n_{gold}(\vec{k})\right)^2 \,  \alpha_0  \,  + \, n_{gold}(\vec{k}) \, (1- \alpha_0 N) \, \right )  + \,  \text{cross-couplings}\,,    
     \label{Goldeffderivatve}
 \end{equation}
   where $n_{gold}(\vec{k}) \, \equiv \, a_{gold}(\vec{k})^\dagger a_{gold}(\vec{k})$  
 is the occupation number of pseudo-Goldstone of a given momentum number $\vec{k}$.  At the critical point, the mass terms vanish and the Hamiltonian is given by the quartic couplings, 
     \begin{equation}
     H_{Gold}  \, = \, \sum_{\vec{k}\neq 0}  \, |\vec{k}|^2  \alpha_0  \left(n_{gold}(\vec{k})\right)^2 \,  \, + \,  \text{cross-couplings}\,.    \label{Goldeffderivatvecrit}
 \end{equation}
We can obtain a lower bound for the number of states in the allowed range by looking for the $|\vec{k}|$ that provides the maximal contribution to the partition sum. We need to take into account two competing effects: The larger $\ak$, the more modes, and thus states, are supplied. On the other hand, the energy cost of a given mode grows with $\ak$, thereby limiting the maximum allowed occupation. Moreover, we pay a penalty for occupying modes more than once as seen from the quadratic dependence of \eqref{Goldeffderivatvecrit} on $n_{gold}(\vec{k})$. A conservative lower bound is thus obtained by summing over all states in a shell around the limiting momentum $k_l$ which allows for all Goldstone modes of given wavenumber to be occupied exactly once. Lower momenta will yield subdominant contributions due to the reduced number of modes and the related penalty for higher occupation; higher momentum modes, on the other hand, are too costly to excite.
%
%
We obtain the contribution, 
       \begin{equation}
     \delta_{k} H_{Gold}  \, \sim  \, |\vec{k_l}|^d \left(|\vec{k_l}|^2  \alpha_0\right) \, . 
         \label{variation}
 \end{equation}
where we ignored a possible cancellation from the cross-couplings.  The factor 
$|\vec{k_l}|^d$ comes from the total number of modes.  The maximal contribution to the entropy 
will come from the species with largest $|\vec{k}|$, subject to the condition  $\delta_{k} H_{Gold} \, < \, 1$. 
Taking into the account that $\alpha_0 \, = \, 1/N$,  and assuming  $L_*/\hbar  =
 m^{-1}$, we obtain
 \begin{equation}
 |\vec{k_l}| = N^{\frac{1}{d+2}}\,\,,\,\,\,n_\text{modes} = N^\frac{d}{d+2}\,.
 \end{equation}
 for the limiting momentum and the corresponding number of modes.
We get the following contribution to the entropy from  the pseudo-Goldstone species,   
      \begin{equation}
      N_{*}  \sim \log{\sum_k^{N^{d/d+2}} {N^{d/d+2} \choose k}} \sim N^{{d \over d+2}} \, . 
         \label{Nstar}
 \end{equation}
Note that we have also evaluated the full partition sum numerically and thereby verified that the wavenumbers with $\ak \approx |\vec{k}_l|$ provide the dominant contributions.
  
   If we take into the account the effect of cross-couplings, ignoring possible "flat directions" 
   on which the different contributions cancel, we get a more conservative lower bound on the entropy. 
    The number of cross-couplings  for a given momentum level $|k|$ scales as the number 
   of corresponding pseudo-Goldstones squared,  
    $\sim  (kd)^{2d}$. Then, equation  (\ref{variation}) changes to
        \begin{equation}
     \delta_{k} H_{Gold}  \, \sim  \, |\vec{k}|^{2d} (|\vec{k}|^2  \alpha_0) \, . 
         \label{variation1}
 \end{equation}
  This lowers the allowed number of simultaneously-excitable Goldstones to 
     \begin{equation}
      N_{*} \ = \,  N^{{d \over 2d+2}} \, . 
         \label{Nstar1}
 \end{equation}
  
   Independently, we observe the general message that  the derivative self-coupling of bosons leads to a
  dramatic increase of pseudo-Goldstone modes.

 \section{The Carriers of  Black Hole Entropy}
 
   We finally wish to address the following question.  Which of  the large number of degenerate states are good candidates for carrying the Bekenstein-type entropy of a black hole? 
 In order to answer this question, we shall distinguish two categories of states.  
 
  {\it 1)}  The first category represents states that 
 {\it cannot} be resolved semi-classically in any macroscopic measurement.  That is, the quantum information stored in such states 
 becomes  unreadable in the infinite-$N$ limit.  We shall refer to this category of states  as type-$A$. 
 
   {\it 2)}  The second category are states that can be resolved in some macroscopic interference experiments. 
 That is, the quantum information stored in such states can be read out even for $N=\infty$.    We shall correspondingly 
 refer to these states as type-$B$. 
   
    What distinguishes these two categories of states  microscopically? 
     In our picture the states of both categories can be labeled by the occupation numbers of some nearly-gapless 
 quantum degrees of freedom.  As discussed above, these information-carriers can be described as Bogolyubov and/or Goldstone degrees of freedom. 
  What distinguishes the two category of states is the scaling behavior of information-carrier occupation numbers in large $N$ limit.  
  
  The type-$A$ category 
of states refers to those in which the relative occupation number of gapless degrees of freedom vanishes 
in the large-$N$ limit. That is,  none of these information-carrier degrees of freedom are macroscopically occupied.  Correspondingly,  the type-$A$ states cannot be resolved in any macroscopic measurements.   

    For type-$B$ states,  some of the Goldstone modes can be macroscopically occupied with an occupation number that scales as a non-vanishing fraction of $N$ in the $N\rightarrow \infty$ limit.  
  Such states can be resolved in macroscopic interference experiments\footnote{In Appendix \ref{sec:appcoherent}, we connect to the classical limit by performing the counting of states in a coherent state basis. In this basis, one can 
	immediately write down the matrix elements that describe macroscopic interference experiments   
	capable of resolving differences  between the different type-$B$ states.}.

 This discussion is more relevant for the derivatively-coupled model  (\ref{Hderivative}), since it exhibits a large diversity of Goldstone modes. However, to keep the discussion as simple as possible, we shall here consider only the model (\ref{Hnonderivative}).  
Therein, let us now exemplify which states can be attributed to the type-A and type-B categories.

As discussed at the end of section \ref{sec:critgs}, at the critical point, two gapless modes emerge.   
     The appearance of these states is a direct consequence of quantum criticality.  
    They come from those components of the $SU(2)$-doublet pseudo-Goldstone that are not affected by the explicit breaking. 
      One of them
      is only an approximate Goldstone, corresponding to one linear superposition of the 
     off-diagonal generators from the quotient $SU(3)/SU(2)\times U(1)$.
   This Goldstone is massless only near the critical point.  
  In contrast, the other degree of freedom
  is an exact Goldstone corresponding to the spontaneously broken 
   $U(1)$-generator $Q=diag(1,0,-1)$ of $SU(3)$.  This Goldstone is massless everywhere at and beyond the critical point. 
   
  A counting of states in the Goldstone as well as in the Bogolyubov language suggests that the number of independent 
 states below the $1/N$ energy gap near the critical point  scales at least as $N^{1/4}$.  
  However, the total number of almost orthogonal states with expectation value of energy below the same gap  becomes larger if we also include coherent states in the counting (see Appendix).  This is because the coherent states are neither energy nor number eigenstates and involve superpositions of arbitrary energetic modes with arbitrarily large occupation numbers.   Correspondingly, they  explore a much bigger fraction of the Hilbert space than the states 
  constructed out of a finite number of energy or number eigenstates.

  Only the states with vanishing fractions of Goldstone (or Bogolyubov) occupation numbers belong to the 
   type-$A$ category.  On the other hand, states with macroscopic occupation numbers of Goldstones, such as 
   coherent states,  belong to the type-$B$ category.  The latter states can be resolved in macroscopic measurements, even in  $N=\infty$ limit.  This is a manifestation of the fact that the coherent states are ``classical".  
   
    
     Given the fact that interference 
    experiments with black holes have never been performed, both of these categories are very interesting, since they both reveal 
    the internal microscopic structure of black holes.  
         However, the question is whether both of the types of states should be counted as carriers of 
     black hole Bekenstein entropy.  
     
       {\it Should  Bekenstein entropy originate exclusively  from the quantum states with small-occupation numbers of many 
   species of gapless Bogolyubov/Goldstone degrees of freedom, or  should the states with large occupation numbers of few species also count? }

   For now, we shall keep both options open. 
   
   \section{Conclusions} 
   
    The purpose of this paper was to emphasize the fundamental role of quantum criticality for  
    information-processing in many-body systems  and to establish this phenomenon as a viable candidate for the underlying 
    mechanism for black hole information-processing, along the lines of \cite{QC, ENT, scrambling}.  
    We showed that the simplest multi-particle systems with critical behavior capture qualitative features of information-processing that is expected for black holes, such as the low energy cost of information storage, 
   large degeneracy of states and scrambling of information.

      In order to visualize the nature of the critical phase transition in the language of spontaneous symmetry breaking, we developed a new description in which we mapped the quantum phase transition in attractive Bose-gas on a Goldstone phenomenon in a sigma model.  The two systems represent two realizations of one and same 
  unitary  symmetry that rotates different momentum modes into each other.  This symmetry is broken both spontaneously as well as explicitly, but the explicit breaking vanishes, up to $1/N$ effects, at the critical point, resulting onto the gapless pseudo-Goldstone modes.               
      
   This mapping allows us to establish a correspondence  between the quantum criticality in Bose-Einstein systems and the change of symmetry breaking pattern in a sigma-model. Correspondingly, we map the  entropy-carrying (nearly)gapless Bogolyubov modes of the Bose-gas on the (nearly)gapless Goldstone modes of the sigma model.  
    This mapping also allows to formulate, up to $1/N$-corrections, an effective theory of nearly gapless modes in the regime in which occupation numbers are large. 
    
     Our findings, within the validity-domain of the description, confirm the results of previous studies 
 \cite{QC, ENT, scrambling} and shed light on the criticality phenomenon from a novel angle. 
 Our studies indicate that the key information-processing properties of black holes are shared by 
 a larger class of the critical systems, including the ones that can be designed in table-top labs. 
 This opens up a possibility of ``borrowing'' black hole information processing  abilities for implementing 
 them in the laboratory systems. 
 
  Since our findings suggest that critical instability is the key for efficient information processing, 
  it would be interesting to find out, by generalizing our approach, whether other unstable systems,  for example, the   
 ones with parametric resonance instabilities \cite{parametric} and large classical statistical fluctuations, studied in \cite{resonance},  also exhibit some analogous  properties of information scrambling and processing.


\section*{Acknowledgements}

 We would like to thank Daniel Flassig and Mischa Panchenko for discussions. 
 G.D. would like to thank the organizers of the ``WE-Heraeus-Seminar: Cold Atoms meet Quantum Field Theory",
 where a part of this work was presented, and co-participants for stimulating discussions and comments. 
 
%

The work of G.D. was supported by the Humboldt Foundation under Alexander von Humboldt Professorship, the ERC Advanced Grant ``UV-completion through Bose-Einstein Condensation (Grant No. 339169) and by the DFG cluster of excellence ``Origin and Structure of the Universe", FPA 2009-07908, CPAN (CSD2007-00042) and HEPHACOSP-ESP00346.
The work of C.G. was supported in part by the Humboldt Foundation and by Grants: FPA 2009-07908, CPAN (CSD2007-00042) and by the ERC Advanced Grant 339169 ``Selfcompletion'' . The work of A.F. was supported by the FCT through the grant SFRH / BD / 77473 / 2011.
The work of N.W. was supported by the Swedish Research Council (VR) through the Oskar Klein Centre.

\appendix

\section{Counting in Coherent State Language}\label{sec:appcoherent}

    In order to highlight the difference between the above two types of states, it is useful to perform the state-counting 
    in a coherent-state basis.  We shall do this on the simplest example of system (\ref{Hamilton}). Generalization to the derivatively-coupled case is straightforward.  
    
    The reason for choosing a coherent-state basis is due to its usefulness  for  establishing the contact with the classical limit.   Since the characteristic property of type-$B$ states is precisely their distinguishability in the classical limit,  the coherent state picture is more convenient for their description.  In particular, in this basis we can 
    	immediately write down the matrix elements that describe macroscopic interference experiments   
    	capable of resolving differences  between the different type-$B$ states. 
    
    Let us evaluate the expectation value of the three-level hamiltonian (\ref{Hamilton}) over a set of coherent states 
    with the fixed total average number $N$.   These states can be labeled by parameter $x$ and the three phases 
    $\theta_0,  \theta_{+1}, \theta_{-1}$ respectively.  We shall denote them by 
    $|x, \theta_0,  \theta_{+1}, \theta_{-1}\rangle$. Each such state represents a tensor product 
    $|x, \theta_0,  \theta_{+1}, \theta_{-1}\rangle \, \equiv \prod_{k=0,\pm1}  \otimes  |N_k\rangle_{coh} $
    of three coherent states, one per each momentum level $k$. These have the following generic form, 
    \begin{equation}
    |N_k\rangle_{coh} \, = {\rm e}^{-{N_{k} \over 2}} \, \sum_0^{\infty} \, {N_{k}^{{n_{k}\over 2}} {\rm e}^{in_k\theta_k} \over \sqrt{n_{k}!}}  |n_{k}\rangle \,,   
    \label{coherent}
    \end{equation}
    where $k =0, -1, +1$.   The mean occupation numbers satisfy  $N_0 = N-2x$ and  $N_{-1} = N_{1} = x$.   
    The integers $n_k \, = \, 0,1,...\infty$ label the Fock space occupation numbers for each of the three levels.  Taking the expectation value 
    of the Hamiltonian and imposing the minimum constraint on the phases (\ref{phaseconstraint}),  we arrive (not surprisingly) at the effective Hamiltonian  (\ref{HamiltonX}).  
    
    Obviously, we should  only count the coherent states that form an almost-orthogonal set.  
    The scalar product of the two distinct coherent states $|\psi \rangle \equiv |x, \theta_0,  \theta_{+1}, \theta_{-1}\rangle$  and  $|\psi' \rangle \equiv |x', \theta_0',  \theta_{+1}', \theta_{-1}'\rangle$
    is, 
    \begin{equation}
    | \langle \psi |\psi'  \rangle|^2 \, = \, {\rm e}^{-  \sum_{k=0,\pm1}  \left ((\sqrt{N_k} - \sqrt{N_k'})^2 + 2\sqrt{N_kN_k'} (1 - 
    	{\rm cos}(\theta_k-\theta_k') )\right) } \,. 
    \label{projection}
    \end{equation} 
    Thus,  the suppression of the matrix element between the two coherent states comes from the two sources: 
    The difference in occupation numbers and difference in phases.  
    
    We shall only be interested in the set of states over which the expectation value of the Hamiltonian 
    stays within $\sim 1/N$ gap. For such states the value of $x$ is bounded from above by $x_{max} \sim \sqrt{N}$. 
    Thus, without taking into the account the variation of phases, 
    we have maximum $\sim N^{1/4}$  coherent states that are orthogonal due to difference  in the occupation number
    $x$.   In addition, there is a degeneracy due to variation of the phases.  For a fixed $x$, we can perform the two independent variations of phases that preserve the minimum constraint 
    (\ref{phaseconstraint}). However, the step of the variation of each independent phase in coherent state counting must be larger than its quantum uncertainty.   From the properties of the coherent states the phases satisfy the quantum uncertainty $\Delta_{quant} \theta_k \sim 1/\sqrt{N_k}$.  Thus, the step in variation of each phase 
    cannot be smaller than this uncertainty.

    For the clarity of  counting we should distinguish variation of the different phases. 
    Notice that since the phases are constrained by the minimum condition   (\ref{phaseconstraint}), there are
    only two independent phase degrees of freedom, which can be chosen as $\theta_0$ and 
    $\bar{\theta} \equiv \theta_{+1}  - \theta_{-1}$.
    The states of interest can thus be labeled as  $|\psi \rangle \equiv |x, \theta_0,  \bar{\theta} \rangle$.

    Let us first fix the phase 
    $\theta_0$ and only vary 
    the phase $\bar{\theta}$.      The quantum uncertainly in this phase is  $\Delta_{quant}\bar{\theta} \sim 1/\sqrt{x}$ and therefore 
    for given  $x$ there are of order $\sqrt{x}$ independent coherent states that are orthogonal to each other.   This set of states is obtained  by variation of the relative phase $\bar{\theta}$ with the elementary step 
    bounded from below by the minimal quantum uncertainty $\sim 1/\sqrt{x}$.    Now summing over different values of $x$ 
    over an interval $0 < x < x_{max}$,  
    we get that, for fixed $\theta_0$, the total number of 
    orthogonal coherent states crowded within the $1/N$ energy gap, is given by  $\sim x_{max} \sim \sqrt{N}$.   
    
        The story with the $\theta_0$ phase is somewhat different. This  phase is a Goldstone boson
         corresponding to the generator $Q_0=diag(1,-2,1)$, which is spontaneously broken already in the 
         under-critical regime $\alpha N \ll 1$. Correspondingly, the coherent states that are orthogonal 
         due to variation of $\theta_0$ exist irrespective of criticality. The number of such orthogonal states for arbitrary value of 
         $x$ is $\sim \sqrt{N-2x}$.  Of course, they continue to  contribute into the degeneracy also at the critical point, 
         but this extra degeneracy has nothing to do with quantum criticality.

\end{document}